\documentclass[12pt]{article}
\usepackage{amsmath}
\usepackage{times}
\usepackage{graphicx}
\usepackage{color}
\usepackage{multirow}
\usepackage[authoryear]{natbib}
\usepackage{rotating}
\usepackage{bbm}
\usepackage{latexsym}

\newcommand{\captionfonts}{\normalsize}

\makeatletter  
\long\def\@makecaption#1#2{%
  \vskip\abovecaptionskip
  \sbox\@tempboxa{{\captionfonts #1: #2}}%
  \ifdim \wd\@tempboxa >\hsize
    {\captionfonts #1: #2\par}
  \else
    \hbox to\hsize{\hfil\box\@tempboxa\hfil}%
  \fi
  \vskip\belowcaptionskip}
\makeatother   

\begin{document}
\hspace{13.9cm}1

\ \vspace{20mm}\\

{\LARGE Advancing System Performance with Redundancy: From Biological to Artificial Designs}

\ \\
{\bf \large Anh Tuan Nguyen$^{\displaystyle 1*}$, Jian Xu$^{\displaystyle 1}$, Diu Khue Luu$^{\displaystyle 1}$, Qi Zhao$^{\displaystyle 2}$, and Zhi Yang$^{\displaystyle 1}$}\\
{$^{\displaystyle 1}$Biomedical Engineering, University of Minnesota, MN, USA.}\\
{$^{\displaystyle 2}$Computer Science and Engineering, University of Minnesota, MN, USA.}\\
{$^{\displaystyle *}$Email: nguy2833@umn.edu.}\\
%

{\bf Keywords:} Representational redundancy, entangled redundancy, precision enhancement, redundant sensing, muscle redundancy, deep residual networks, bio-inspired designs

\thispagestyle{empty}
\markboth{}{NC instructions}
\ \vspace{-0mm}\\
%
\begin{center} {\bf Abstract} \end{center}
Redundancy is a fundamental characteristic of many biological processes such as those in the genetic, visual, muscular and nervous system; yet its function has not been fully understood. The conventional interpretation of redundancy is that it serves as a fault-tolerance mechanism, which leads to redundancy's de facto application in man-made systems for reliability enhancement. On the contrary, our previous works have demonstrated an example where redundancy can be engineered solely for enhancing other aspects of the system, namely accuracy and precision. This design was inspired by the binocular structure of the human vision which we believe may share a similar operation. In this paper, we present a unified theory describing how such utilization of redundancy is feasible through two complementary mechanisms: \textit{representational redundancy} (RPR) and \textit{entangled redundancy} (ETR). Besides the previous works, we point out two additional examples where our new understanding of redundancy can be applied to justify a system's superior performance. One is the human musculoskeletal system (HMS) - a biological instance, and one is the deep residual neural network (ResNet) - an artificial counterpart. We envision that our theory would provide a framework for the future development of bio-inspired redundant artificial systems as well as assist the studies of the fundamental mechanisms governing various biological processes.

\section{Introduction}
Redundancy is a well-known characteristic of many biological processes from the molecular to the systematic level. For example, the human's genome is highly redundant: a particular gene can be duplicated at various regions of DNA while multiple genes can encode the same or similar biochemical functions and phenotype expressions. These genetic redundancy and functional redundancy are observed in many crucial pathways of the developmental, signaling, and cell cycle processes \citep{1992_Tautz, 1997_Nowak, 2009_Kafri}. High level of redundancy can also be found in the nervous system. The neuronal architecture and synaptic interconnections have been shown to be highly redundant which allows them to facilitate complex processes of information processing, learning, memorizing, and self-repairing. In fact, it is believed that the human brain is at least twice the size as necessary for its function as a result of neural redundancy \citep{1987_Glassman}. 

In many scenarios, the redundant structure of a biological system can be seen as a consequence of the evolutionary process. Under the pressure of natural selection, living organisms develop multiple different strategies that achieve the same goal: survival. It is not uncommon for distinct strategies that emerge from entirely different evolutionary pathways to resolve the same biological problem. These strategies could co-exist in the same ecosystem or even the same organism's genome creating observable repeated evolutional behaviors such as functional redundancy, parallel evolution, and convergent evolution \citep{2017_York}. Redundancy also serves as a defence mechanism against failures which contribute to a higher survival rate. For example, gene duplication has been shown to mitigate effects of mutations and reduce the chance of catastrophic phenotype expression \citep{2009_Kafri}. Redundancy also helps the human brain tolerate significant damages and loss of mass due to injuries or diseases. Damaged neurons and brain tissue generally do not regrow, yet their redundant structures allow reorganization of the neuronal circuits to recover many basic brain functions \citep{1987_Glassman}. Lastly, redundancy increases the organism's adaptivity. For example, genetic and functional duplication has been shown to be the basis of phenotypic plasticity which allows an organism to adapt and survive rapidly-changing endogenous and exogenous environmental conditions \citep{2009_Kafri}.

Many of these principles find their application in designing and engineering of artificial systems. However, almost all intentional utilization of redundancy in man-made systems focus on enhancing reliability, which importance is often overshadowed by the system's performance. Also, existing methods for incorporating redundancy involves the replication of partial or entire systems which require large resources overhead. As a result, redundant designs such as dual modular redundancy (DMR) or triple modular redundancy (TMR) are mostly found in specialized systems that perform critical functions such as aircraft controllers, biomedical implants, and computer servers, etc.

In this paper, we argue two counter-intuitive arguments. First, redundancy can be engineered solely for enhancing systems' performance regarding accuracy and precision, instead of reliability and plasticity. Secondly, a practical implementation of redundancy is feasible without replication and excessive resource overhead, thus mitigating the trade-off encountered by conventional designs. The performance boost in our proposed framework is achieved by employing two complementary mechanisms, namely RPR and ETR. RPR describes how information is redundantly encoded and processed, while ETR allows realizing of a RPR scheme in actual applications.

In \citep{2015_Nguyen, 2016_Nguyen_NIPS}, we have shown a simple but practical application where redundancy resembling the binocular structure of the human vision is applied to enhance the precision of a man-made sensor without incurring compromises often seen in conventional architectures. In theory, the RPR and ETR principles utilized in our design can be generalized to different applications, and also serve as a fundamentally structural characteristic of more complex systems. This argument is further asserted in this paper by examining empirical evidence in two different systems from two distinct fields of science and engineering. One is the HMS - a biological system where redundancy contributes to generating complex and precise muscle movements; another is the ResNet - an artificial deep learning architecture where redundancy helps accomplish superior predicting accuracy compared to conventional methods. By understanding subtle yet sophisticated roles of redundancy in these systems, we believe that the findings would not only enrich our knowledge of biological processes but also inform the derivation of new methods for advancing the performance of man-made designs. 

The remains of the paper are organized as follows. Section 2 consolidates our redundant model comprised of RPR and ETR mechanism. Section 3 examines the evidence suggesting the implication of our model in biological and artificial systems, which include the proposed sensor design, the HMS, and the ResNet. Finally, section 4 concludes our findings and gives discussions on the future development of the proposed theory.

\section{Advancing Performance with Redundancy}

\textbf{Representational redundancy:} The vast majority of artificial systems are designed upon an orthogonal scheme of information representation where each entry of information is encoded by a unique configuration of the system. An entry of information can be an input value, a desirable output, an intermediate instance or an operation of the information processing pathway. Such orthogonal systems excel in efficiency because they allow rapidly and unambiguously acquiring, processing and storing of information. However, any encoding/decoding scheme in practice suffers from an inevitable level of error resulting in the limitation of its accuracy. In many computational models, this limitation is described by Shannon's theorem. Because of the uniqueness of the representation scheme, any error acquired during the sampling, processing and storing of information cannot be easily corrected without an overhead in term of resources such as power, bandwidth, and memory, etc.

The RPR concept is designed to overcome conventional limitations by embracing a non-orthogonal scheme of information representation. Subsequently, every entry of information can be encoded by numerous distinct system configurations, including the conventional one. These configurations are referred as the system's \textit{microstates}. If the microstates are designed such that their response to error are non-homologous, in any given instance, provided a sufficient number of distinct microstates, there exist with asymptotic certainty one or more microstates that have a smaller error than the conventional representation. Therefore, an overall RPR-system would have a theoretical accuracy almost always superior to the conventional counterpart with similar structure. 

\textbf{Entangled redundancy:} The number of microstates represents the information capacity - an abstract property of the design that is not necessarily proportional to its physical size. In order to effectively deploy a RPR-system in practice, the microstates must be designed so that they do not incur excessive resource overhead. As a result, the statistical distribution of the microstates with respect to error cannot be independent, but partially correlated or \textit{entangled}. This concept is known as ETR. The level of entanglement should be engineered just sufficient to create excessive redundancy without trading off large amounts of resource. ETR should be differentiated from the conventional method of creating redundancy by replication where the distribution of repeated instances are independent of each other and the resource utilization is linearly proportional to the level of redundancy. 
\begin{figure}[!h]
\begin{center}
\includegraphics[width=1\textwidth]{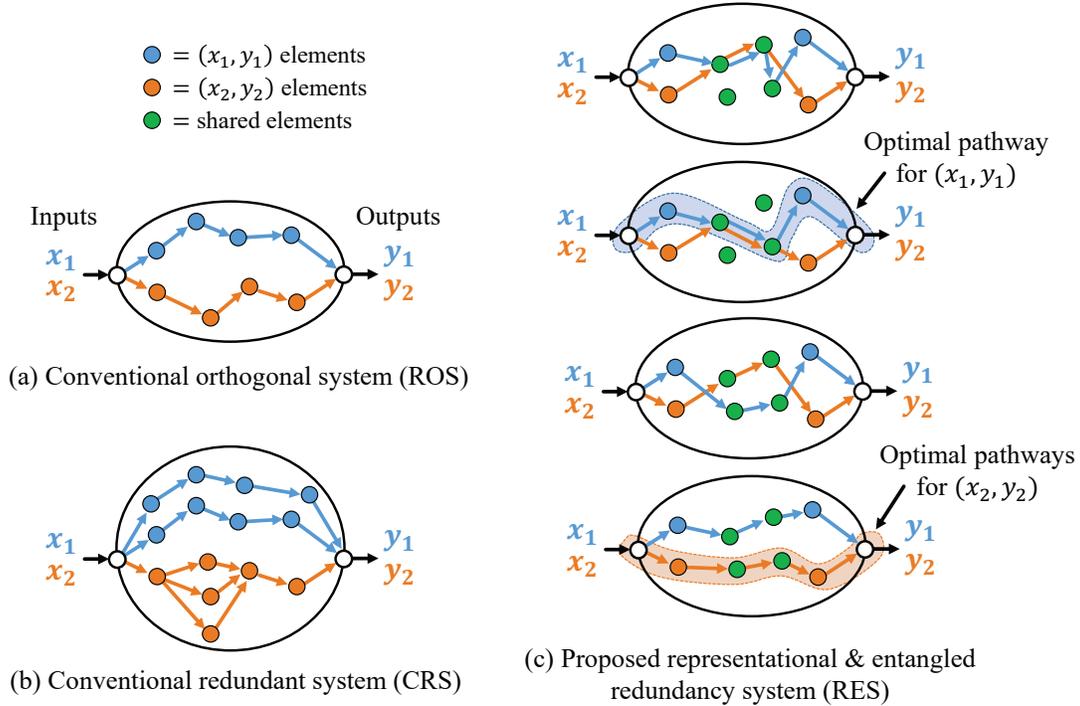}
\end{center}
\vspace{-5mm}
\caption{Illustration of the differences among a conventional orthogonal system (COS), a conventional redundant system (CRS) and a proposed RPR+ETR system (RES). An entry of information is a processing pathway that takes an input $x_i$ and produce a corresponding output $y_i$ ($i = 1,2,...$). (a) Every input/output (I/O) pair in the COS is represented by a unique pathway. (b) The pathways in the CRS are partially or entirely replicated which give the system fault-tolerance properties and a marginal accuracy gain. (c) The proposed RES achieves redundancy by having different pathways share certain elements. The entanglement allows exponential level of redundancy with minimal additional resources. For each I/O pair, the pathway with the least error can be selected resulting in major accuracy enhancement.}
\label{Fig_RedunInfo}
\end{figure}

Figure \ref{Fig_RedunInfo} illustrates the distinction between a conventional orthogonal system (COS), a conventional redundant system (CRS) and a proposed RPR+ETR system (RES). An entry of information in this illustration is a processing pathway that takes an input $x_i$ and produce a corresponding output $y_i$ ($i = 1,2,...$). In the COS, every input/output (I/O) pair $(x_i, y_i)$ is represented by a unique pathway which has a determined error that cannot be easily removed. The pathways in the CRS are partially or entirely replicated, which requires a proportional resource overhead. Although in practice, the replication is mostly used for fault-tolerance, a marginal accuracy gain is feasible by selecting the pathway with the least error for each input instance. The RES incorporates redundancy by having the pathways of different I/O pairs share certain elements. Each $(x_i, y_i)$ pathways can now be represented by multiple system's pathways, i.e. microstates, which number increases exponentially with the number of shared elements. The RES is superior compared to COS because there almost always exists a pathway with a lower error for any given I/O pair. The RES is also superior compared to CRS because an exponential level of redundancy can be achieved with minimal additional resources.

\textbf{Challenges:} A proper implementation of RPR and ETR in the same architecture is essential to achieve the performance boost. The goal is to create an excessive number of microstates while utilizing their entanglement to allow the microstates to co-exist in superposition thus requiring minimal additional resources as shown in Figure \ref{Fig_RedunInfo}(c). Unfortunately, there is no universal solution that can be applied to all types of system. In our proof-of-concept system described in \citep{2015_Nguyen, 2016_Nguyen_NIPS}, redundancy is realized by integrating two similar binary-weighted arrays, which structure is speculated to resemble the human's binocular vision. In the subsequent sections, the HMS and ResNet provide additional examples where redundancy elegantly emerges in entirely different manners.

Furthermore, provided a redundant non-orthogonal structure of information representation, there is no universal solution to identify the optimal microstate given a particular input. In fact, in almost all examples of RES, it appears to be an NP-optimization problem that can only be resolved by the mean of approximation. Biological processes such as the visual and musculoskeletal system overcome this challenge by harnessing the computational capacity of the nervous system, which is exceptionally adequate at approximation. A similar mechanism could be utilized by the ResNet which itself is a neural network. For engineering systems such as our redundant sensor \citep{2015_Nguyen}, an approximation method need to be derived, which consists of a one-shot unsupervised error estimation and a simplified calibration algorithm.

\section{From Biological to Artificial Systems}

\subsection{Redundant Sensing}
\begin{figure}[!h]
\begin{center}
\includegraphics[width=1\textwidth]{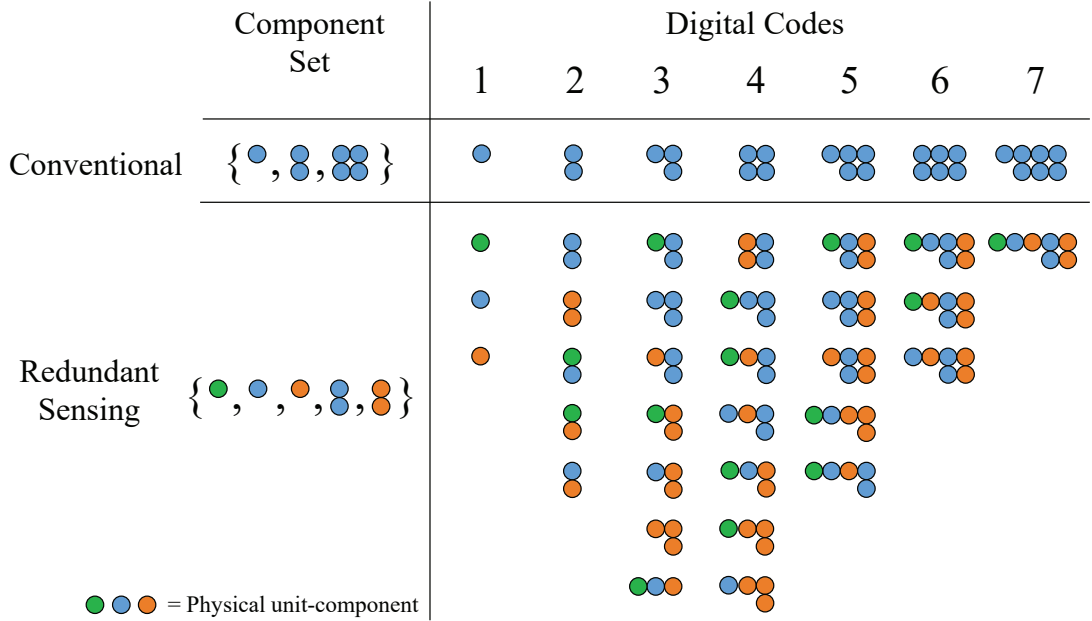}	
\end{center}
\vspace{-5mm}
\caption{Illustration of the RPR and ETS properties of the proposed redundant sensing (RS) architecture \citep{2015_Nguyen}. The device converts an analog input to a digital output by assembling a set of physical components. While utilizing the same number of unit-components ($2^N-1=7$, $N$: sensor's resolution) as a conventional design, the component set of a RS architecture allows each digital code to be created by multiple different assemblies, i.e. microstates. By selecting the microstate with the least error for every code, a significant boost of the accuracy can be achieved. This example illustrates a simplified case of $N=3$. The RS is most effective for high resolution because the number of microstates increases exponentially with $N$.}
\label{Fig_RedunSense}
\end{figure}

The work of \citep{2015_Nguyen} shows a proof-of-concept implementation of a RPR+ETR system: a sensor that converts analog to digital signals. A entry of information is a digital code $x_D \in \{0,1,..., 2^N-1\}$ ($N = $ resolution) representing an input analog voltage. In practice, each code is generated by assembling a set of components which are miniature capacitors embedded on a silicon chip which number is proportional to the required physical resources and cost. The random error occurred during the fabrication process of these capacitors, i. e. mismatch error, has been shown to be a major factor limiting the device's accuracy. 

Figure \ref{Fig_RedunSense} compares the differences between a conventional and a RS architecture in a simplified case of $N=3$. The conventional system utilizes a binary-weighted set of components which is the most efficient encoding scheme yet vulnerable to mismatch error. On the other hand, the proposed RS architecture employs a non-orthogonal component set which satisfies both RPR and ETR requirements. With the name number of unit-component ($2^N-1=7$), RS allows each digital code to be generated by multiple different component assemblies, i.e. microstates. The number of microstates increases exponentially with $N$ resulting in major accuracy enhancement of high-resolution devices.

Interestingly, the RS component set resembles exchanging and integrating the information between two smaller conventional binary-weighted sub-arrays, which is inspired by the binocular structure of the human visual system. Thus, we ask the question: whether RPR and ETR are fundamental properties that facilitate visual acuity? The spatial distribution of photoreceptors on the retina is notably irregular, which echoes the impact of mismatch error. It is therefore conjectured that binocular vision effectively creates a form of static redundancy allowing the brain to collect sufficient information to remedy the error and produce the images we perceive. In fact, the binocular vision has been shown to help differentiation of fine details, even exceeding the diffraction limit of the photoreceptors - a phenomenon known as hyperacuity \citep{1979_Beck}. Our theory is further supported by the fact that human and many higher-order animals only have two eyes. During the development of our RS sensor, we found out that the amount of computational power required to process redundant information increases rapidly with the number of sub-arrays. Two sub-arrays or two eyes is the minimum number necessary to create a redundant structure. Despite dedicating 30-60\% of its mass for visual processing, the brain simply lacks the capacity to process information from three or more eyes.

Furthermore, as a complement to the binocular structure, we conjecture that eyes' micro-fixational movement or microsaccade \citep{2013_MartinezConde} creates a form of dynamic redundancy. During microsaccades, the field of vision of each eye is sampled multiple times by different spatial configurations of photoreceptors, which resemble entangled redundant microstates and facilitate visual acuity. This observation is supported by both experiments with human subjects \citep{2013_Hicheur} and mathematical modeling \citep{2004_Hennig} where microsaccades have been shown to play an important role in the visual precision and could lead to hyperacuity.

\subsection{Muscle Redundancy}
\begin{figure}[!h]
\begin{center}
\includegraphics[width=0.9\textwidth]{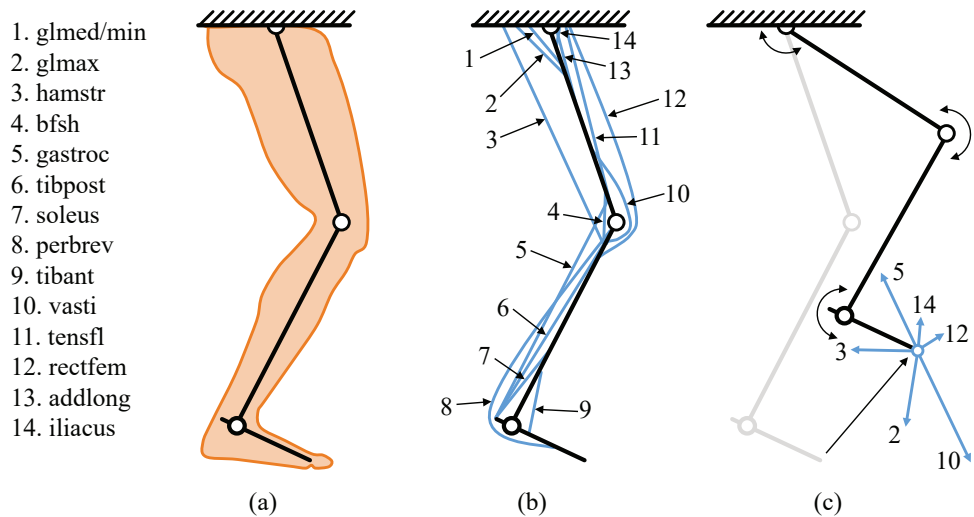}
\end{center}
\vspace{-5mm}	
\caption{(a, b) A sagittal view of a human leg's mechanical model consisting of 14 muscles and muscle groups \citep{2011_Kutch}. (c) Any specific movement trajectory and force can be achieved by multiple distinct muscles and joints combinations. Also, different muscles have overlapping but not exclusive mechanical functions, and all contribute with different degrees in generating force and movement. These characteristics resemble a RPR+ETR system.}
\label{Fig_MuscleRedun}
\end{figure}

The HMS has more muscles and joints than the necessary mechanical degrees-of-freedom even though are energetically expensive to produce and maintain. This paradoxical phenomenon of muscle redundancy (MS) presents a long-standing problem in human kinesiology of understanding how and why the human brain coordinates all muscles and joints to achieve complex movements with precision \citep{1967_Bernstein}. By examining this biological process from the perspective of our model, we hope to unravel the principles underlying the behavior of MS. 

A conventional interpretation would suggest that redundancy contributes to the reliability of the HMS allowing compensation for the loss or dysfunction of individual muscles. However, emerging empirical evidence suggests this is not true. Even a mild dysfunction of a few critical muscles due to disorders, injuries or aging can significantly weaken the force production and overall functions of the whole HMS \citep{1991_Forssberg, 2006_Schreuders}. The results are supported by Kutch \& Valero-Cuevas works \citep{2011_Kutch, 2015_ValeroCuevas}. Using both computational models and empirical experiments with cadaver specimens, the authors point out that less than 5\% of the feasible forces and movements in their models are robust to a loss of any muscle, so it is clear that reliability is not an inherited characteristic of MS.

Figure \ref{Fig_MuscleRedun} presents a sagittal view of a human leg's mechanical model used by Kutch \& Valero-Cuevas \citep{2011_Kutch} which consists of 14 muscles and muscle groups\footnote{List of 14 muscles/muscle groups and their abbreviation: (1) gluteus medialis and minimus (glmed/min); (2) gluteus maximus (glmax); (3) semimembranoseus, semitendenosis and biceps femoris long head (hamstr); (4) biceps femoris short head (bfsh); (5) medial and lateral gastrocnemius (gastroc); (6) tibialis posterior (tibpost); (7) soleus (soleus); (8) peroneus brevis (perbrev); (9) tibialis anterior (tibant); (10) vastus intermedius, lateralis and medialis (vasti); (11) tensor facia lata (tensfl); (12) rectus femoris (rectfem); (13) adductor longus (addlong); (14) iliacus (iliacus).}. At the kinematic and muscular level, any specific movement trajectory and force can be achieved by virtually infinite combinations of muscles and joints. While at the control level, each muscle consists of numerous units that can be activated by different motor neurons and patterns while resulting in the same behavior. Furthermore, the muscles have overlapping but not exclusive mechanical functions, and all contribute with different degrees in generating force and movement. Clearly, these characteristics of the HMS resemble ones of a RPR+ETR system, thus our model predicts that MR plays a major role in enhancing the accuracy and precision of muscle movements.

Indeed, the hypothesis is supported by a number of studies. In \citep{2010_Cleather}, the authors examine two different muscular models, namely Delp and Horsman, in predicting the patellofemoral force during standing, jumping, and weightlifting. They conclude that higher level of redundancy in the Horsman model contributes to its higher predictive accuracy and closer realistic approximation in all activities. The authors' conjecture is consistent with our theory which implies redundancy effectively increases the variability and number of independent musculoskeletal movements (i.e. microstates), so an optimal solution is more likely to be found. The argument is further strengthened in \citep{2016_Moissenet} where an increased level of redundancy correlates to a better predicting accuracy of tibiofemoral contact forces in all gait patterns.

\subsection{Deep Residual Networks}
\begin{figure}[!h]
\begin{center}
\includegraphics[width=0.8\textwidth]{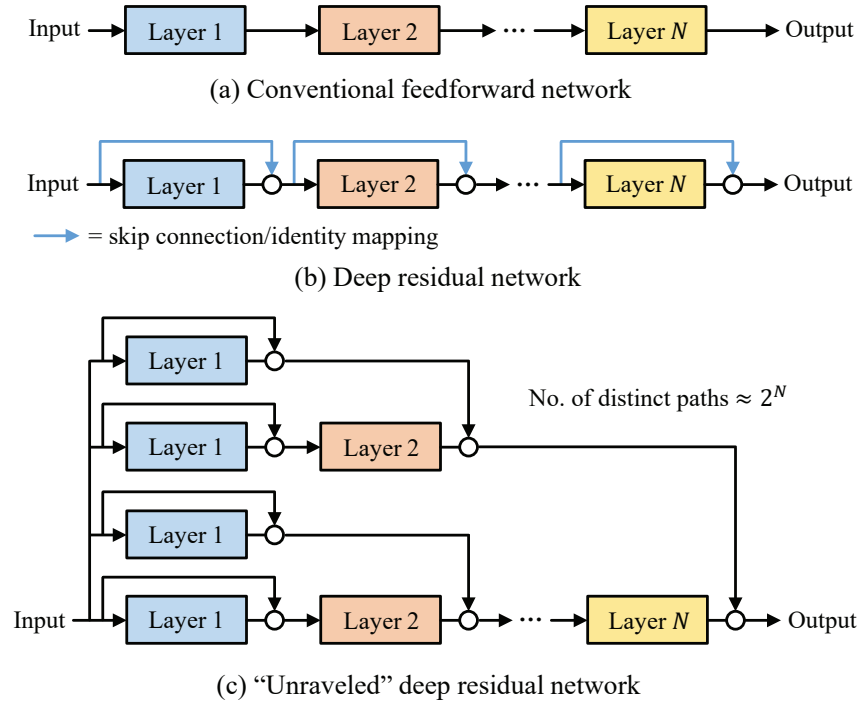}	
\end{center}
\vspace{-5mm}
\caption{(a) A conventional feedforward DNN produces a prediction by convoluting the input through multiple feature layers. (b) A ResNet achieves superior predictive accuracy with the incorporation of skip connections/identity mappings which allow information to bypass an entire layer. (c) It has been shown that the behavior of the ResNet is similar to a collection of shallower networks, resembling a RPR+ETR system \citep{2016_Veit}.}
\label{Fig_ResNet}
\end{figure}

There exist several prominent uses of redundancy for enhancing classification accuracy in machine learning. In a ``committee machine'' or ``ensemble learning'', multiple predictions are generated simultaneously by a collection of discrete instances that are based on the same or distinct predictive models. Because each instance produces a result with a different degree of error, an appropriate integration of these outcomes could lead to a higher overall accuracy \citep{2006_Bishop}. Another approach involves replication of individual neurons or sub-circuits of an artificial neural network (ANN). Using mathematical models, \citep{1990_Izui, 1988_Tanaka} conclude that replication can fundamentally alter the computation carried out by an ANN resulting in quantitative enhancement of convergence speed, solution accuracy, interconnection stability. The findings were utilized to design redundant ANNs simulating a robotic arm grasping an object in 2D space and a pattern-classification task with improved accuracy and convergence time \citep{1994_Medler_1, 1994_Medler_2}. These are prime examples of CRS as shown in Figure \ref{Fig_RedunInfo}(b). Even though a marginal accuracy boost can be accomplished, the replication-based implementation prevents these systems from effectively employing redundancy without incurring excessive resource overhead.

Recently, deep learning has emerged as a leading field of machine learning \citep{2015_LeCun}. A feedforward deep neural network (DNN) produces a prediction by convoluting the inputs through various feature layers encoding the acquired knowledge (Figure \ref{Fig_ResNet}(a)). One of the breakthroughs in DNN design - the ResNet \citep{2016_He_1, 2016_He_2} - modifies the conventional structure by including ``skip connections'' or ``identity mapping'' that allow information to occasionally bypass an entire layer (Figure \ref{Fig_ResNet}(b)). Empirical experiments have demonstrated the superior predictive accuracy of ResNet compared to conventional networks with the same number of layers and parameters \citep{2016_He_1, 2016_He_2, 2016_Huang, 2016_Wu, 2017_Zagoruyko}.

Although the advantage of ResNet is evident, many are baffled by how a subtle yet critical modification of the DNN could fundamentally alter its properties. It becomes clear as Veit \textit{et al.} \citep{2016_Veit} show that the ResNet's behaviors resemble characteristics of an ensemble of shallower networks. As illustrated in Figure \ref{Fig_ResNet}(c), the network can be ``unraveled'' as a sum of smaller sub-circuits where information can flow through any one of the $2^N$ distinct pathways ($N$: number of layers) and is integrated at the last step. The structure resembles a proposed RPR+ETR system where each of pathway corresponds to a microstate. Because of the entanglement among microstates, an excessive level of redundancy which is exponentially proportional to the number of layers can be formulated without compromising the size of the network. 
Therefore, we argue that by including the skip connections, the conventional DNN has been transformed into a redundant system with both RPR and ETR properties which leads to major enhancement of performance.

\section*{Discussion \& Conclusion}

Although redundancy is no doubt an essential property of many biological processes, there are reasons to believe that its functions have not been fully appreciated resulting in the absence in artificial designs. While the conventional interpretation ties redundancy with fault-tolerance, we propose a new model arguing that it can be engineered to advance the performance regarding accuracy and precision. Our theory highlights two fundamental mechanisms enabling such function: (i) RPR facilitates redundant encoding of information, and (ii) ETR facilitates practical implementation of redundancy. Besides suggesting the presence of these mechanisms in biological processes such as the human visual and musculoskeletal systems, we present two state-of-the-art man-made designs, the RS sensor \citep{2016_Nguyen_NIPS} and the ResNet \citep{2016_He_1}, where redundancy is successfully employed.  

Clearly, there are future works needed to be done to demonstrate the feasibility of such redundant architectures. Firstly, under the guidelines of our framework, new engineering solutions should be derived to integrate redundancy into other designs for accuracy and precision enhancement. Although the principles of RPR and ETR are universal, their actual implementations vary drastically. The examples in electrical engineering and computer pointed out in this paper are merely the tip of the iceberg. Secondly, a new technique should be investigated to evaluate the information capacity of redundant systems which correlates to its upper bound of performance. A brute force approach utilized in the RS design \citep{2016_Nguyen_NIPS} certainly cannot be applied to more complex systems such as the ResNet. Finally, new methods should be developed to harness the full capacity of redundant systems. Redundant representation of information is irrelevant without an effective way to extract the optimal configuration. In almost all examples shown in this work, they present NP-optimization problems which solutions can be adequately obtained by the mean of approximation.



\end{document}